\documentclass[pra,showpacs,showkeys,amsfonts,amsmath,twocolumn]{revtex4}
\usepackage{bm}
\usepackage{graphicx}
\RequirePackage{mathptm}

\numberwithin{equation}{section}

\newcommand{\sa}[2]{\sigma_{#1}^{#2}}

\newcommand{\ro}{\rho}
\newcommand{\ras}{\rho_{\mr{as}}}

\newcommand{\al}{\alpha}

\newcommand{\g}{\gamma}
\newcommand{\ga}[1]{\gamma_{#1}}
\newcommand{\Ga}[1]{\Gamma_{#1}}
\newcommand{\Om}[1]{\Omega_{#1}}

\newcommand{\conj}[1]{\overline{#1}}
\newcommand{\ket}[1]{|{#1}\rangle}

\newcommand{\C}{\mathbb C}

\newcommand{\mr}[1]{\mathrm{#1}}

\begin{document}
\title{Dynamical creation of entanglement versus disentanglement in a system
of three - level atoms with vacuum - induced coherences}
\author{{\L}ukasz Derkacz}
\affiliation{Institute of Theoretical Physics\\ University of
Wroc{\l}aw\\
Plac Maxa Borna 9, 50-204 Wroc{\l}aw, Poland}
\author{Lech Jak{\'o}bczyk
\footnote{E-mail addres: ljak@ift.uni.wroc.pl} }
 \affiliation{Institute of Theoretical Physics\\ University of
Wroc{\l}aw\\
Plac Maxa Borna 9, 50-204 Wroc{\l}aw, Poland}
\begin{abstract}
The dynamics of entanglement between  three - level atoms coupled to
the common vacuum is investigated. We show that the collective
effects such as collective damping, dipole - dipole interaction and
the cross coupling between orthogonal dipoles, play a crucial role
in the process of creation of entanglement. In particular, the
additional cross coupling enhances the production of entanglement.
For the specific initial states we  find that the effect of delayed
sudden birth of entanglement, recently invented by Ficek and Tana\'s
[Phys. Rev. A \textbf{77}, 054301(2008)] in the case of two - level
atoms, can also be observed in the system. When the initial state is
entangled, the process of spontaneous emission causes destruction of
correlations and its disentanglement. We show that the robustness of
initial entanglement against the noise can be changed by local
operations performed on the state.
\end{abstract}
\maketitle
\section{Introduction}
In a system of coupled multi - level atoms having closely lying
energy states and interacting with the vacuum, quantum interference
between different radiative transitions can occur, resulting in
coherences in a system which are known as \textit{vacuum - induced
coherences}. For example, when the distance between  atoms is
comparable to the wavelength of the emitted radiation and transition
dipole moments involving the decay processes are parallel, the
coupling between the atoms via common vacuum gives rise to the
collective effects such as collective damping and dipole - dipole
interaction. Such effects are well known \cite{Ag, FS}, particularly
in the case of two - level atoms. In the system of three - level
atoms, radiative coupling can produce a new interference effect in
the spontaneous emission. This effect manifests by the cross
coupling between radiative transitions with \textit{orthogonal}
dipole moments \cite{AP} and is strongly dependent on the relative
orientation of the atoms \cite{EK,SE}. All such collective
properties of the system influence the quantum dynamics, which can
significantly differ from a corresponding single atom dynamics.
There were many studies on the effect of quantum interference on
various physical processes including: resonance fluorescence
\cite{HP}, quantum jumps \cite{Z}, the presence of ultranarrow
spectral lines \cite{ZS} or amplification without population
inversion \cite{Ha}.
\par
In our research we consider entanglement properties of a pair of
three - level atoms in the \textsf{V} configuration with vacuum
induced coherences. We study a dynamical creation of entanglement
due to the collective effects  which are present in the system,  as
well as the process of degradation of correlations, resulting in
disentanglement of initially entangled pairs of atoms. Both
processes crucially depend on the interatomic distance compared to
the wavelength of the emitted radiation. For large separation we
expect that the collective properties of two atoms are negligible
and dissipation causes disentanglement. On  the contrary, for small
distance the collective effects are so strong that they can
partially overcome decoherence. As a result, the system can decay to
a stationary state which can be entangled, even if the initial state
was separable \cite{DJ1}.
\par
In the present paper, we study the case of distance comparable to
the radiation wavelength $\lambda$. Although the dynamics brings all
initial states into the asymptotic state in which both atoms are in
their ground states, still there can be some transient entanglement
between the atoms. In particular we show that the dynamical creation
of entanglement is possible in a system where only one atom is in
the excited state. Moreover the production of entanglement is
enhanced, when the cross coupling between orthogonal dipoles is
present. In the more accessible initial state when the both atoms
are excited, and if the cross coupling is absent, the interesting
phenomenon of delayed sudden birth of entanglement \cite{FT} can be
observed: unentangled atoms become entangled after some finite time,
despite of the fact that the correlation between the atoms existed
earlier. On the other hand, cross coupling causes that the
entanglement starts to build up immediately. We consider also the
process of disentanglement of initially entangled states in the
presence of vacuum induced coherences. Analogously to the case of
two - level atoms (see e.g. \cite{FTpr}), there are specific
entangled states of our system which decay much slower then the
other states. In the limit of small separation, those states
decouple from the environment and therefore are stable. They are
called (generalized) antisymmetric Dicke states \cite{BGZ} and play
the crucial role in characterizing disentanglement properties of
given initial state. In particular, the class of maximally entangled
states of two - qutrits i.e. (generalized) Bell states \cite{BHN}
can be divided into two subsets. The first set contains those states
which have no populations in antisymmetric Dicke states, and they
decay rapidly. The remaining states have equal populations in stable
Dicke states and decay much slower. Since all Bell states are
locally equivalent,  local operations performed on the states  may
change the robustness of entanglement against the noise.
\section{Model dynamics}
We start with a short description of the model studied by Agarwal
and Patnaik \cite{AP}. Consider two identical three - level atoms (
$A$ and $B$) in the $V$ configuration. The atoms have two near -
degenerate excited states $\ket{1_{\al}},\; \ket{2_{\al}}$ ($\al
=A,B$) and ground states $\ket{3_{\al}}$. Assume that the atoms
interact with the common vacuum and that transition dipole moments
of atom $A$ are parallel to the transition dipole moments of atom
$B$. Due to this interaction, the process of spontaneous emission
from two excited levels to the ground state  take place in each
individual atom but a direct transition between excited levels is
not possible. Moreover, the coupling between two atoms can be
produced by the exchange of the photons. As it was shown by Agarwal
and Patnaik, in such atomic system there is also possible the
radiative process in which atom $A$ in the excited state
$\ket{1_{A}}$ loses its excitation which in turn excites atom $B$ to
the state $\ket{2_{B}}$. This effect manifests by the cross coupling
between radiation transitions with orthogonal dipole moments. The
evolution this atomic system can be described by the following
master equation \cite{AP}
\begin{equation}
\frac{d\ro}{dt}=(L^{A}+L^{B}+L^{AB})\ro\label{me}
\end{equation}
where for $\al=A,B$ we have
\begin{equation}
L^{\al}\ro=\sum\limits_{k=1}^{2}\ga{k3}\,\left(\,2\sa{3k}{\al}\ro\sa{k3}{\al}-\sa{a3}{\al}
\sa{3k}{\al}\ro-\ro\sa{k3}{\al}\sa{3k}{\al}\right)  \label{genA}
\end{equation}
and
\begin{equation}
\begin{split}
L^{AB}\ro=&\hspace*{3mm}\sum\limits_{k=1}^{2}\sum\limits_{\al=A,B}\Ga{k3}\,(\,
2\sa{3k}{\al}\ro\sa{k3}{\neg\al}-\sa{k3}{\neg\al}\sa{3k}{\al}\ro-\ro\sa{k3}{\neg\al}\sa{3k}{\al})\\[1mm]
&\hspace*{3mm}+i\,\sum\limits_{k=1}^{2}\Om{k3}\,
\left[\,\sa{k3}{A}\sa{3k}{B}+\sa{k3}{B}\sa{3k}{A},\ro\,\right]\\[1mm]
&\hspace*{3mm}+\Ga{vc}\sum\limits_{\al=A,B}(\,2\sa{31}{\al}\ro\sa{23}{\neg\al}
-\sa{23}{\neg\al}\sa{31}{\al}\ro-
\ro\sa{23}{\neg\al}\sa{31}{\al}\\[1mm]
&\hspace*{20mm}+2\sa{32}{\al}\ro\sa{13}{\neg\al}-\sa{13}{\neg\al}\sa{32}{\al}\ro
-\ro\sa{13}{\neg\al}\sa{32}{\al}\,)\\[2mm]
&\hspace*{3mm}+i\,\Om{vc}\sum\limits_{\al=A,B}
[\sa{23}{\al}\sa{31}{\neg\al}+\sa{32}{\al}\sa{13}{\neg\al},\ro\,]
\label{genAB}
\end{split}
\end{equation}
In the equations (\ref{genA}) and (\ref{genAB}), $\neg\al$ is $A$
for $\al=B$ and $B$ for $\al=A$, $\sa{jk}{\al}$ is the transition
operator from $\ket{k_{\al}}$ to $\ket{j_{\al}}$  and the
coefficient $\ga{j3}$ represents the single atom spontaneous - decay
rate from the state $\ket{j}$ ( $j=1,2$ ) to the state $\ket{3}$.
The coefficients $\Ga{j3}$ and $\Om{j3}$ are related to the coupling
between two atoms and are the collective damping  and the dipole -
dipole interaction potential, respectively. The coherence terms
$\Ga{vc}$ and $\Om{vc}$ are cross coupling coefficients, which
couple a pair of orthogonal dipoles. This cross coupling between two
atoms strongly depend on the relative orientation of the atoms. To
see this, we put the atom A at the origin of coordinate system and
the position of the atom B is give by the vector $\vec{R}$ which
makes na angle $\phi$ with the $x$ axis and an angle $\theta$ with
the $z$ axis (see FIG. 1). Assume that the dipole moments
$\vec{d}_{13}$ and $\vec{d}_{23}$ of transitions $\ket{1_{\al}}\to
\ket{3_{\al}}$ and $\ket{2_{\al}}\to \ket{3_{\al}}$ are given by
$$
\vec{d}_{13}=\hat{x}d,\quad \vec{d}_{23}=\hat{y}d
$$
\begin{figure}[t]
\centering {\includegraphics[height=52mm]{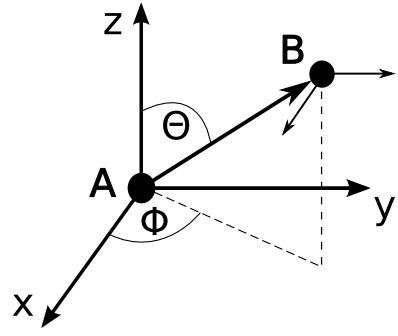}}\caption{The
considered geometry of two - atomic system.}
\end{figure}
Since the states $\ket{1_{\al}}$ and $\ket{2_{\al}}$ are closely
lying, the transition frequencies $\omega_{13}$ and $\omega_{23}$
satisfy
$$
\omega_{13}\approx\omega_{23}=\omega_{0}
$$
Similarly, the spontaneous - decay rates
$$
\ga{13}\approx\ga{23}=\g
$$
As was shown in Ref.\cite{AP}, the coefficients in (\ref{genAB}) can
be written as
\begin{equation}
\begin{split}
&\Ga{13}=\frac{3\g}{2}\left(P_{i}-\sin^{2}\theta\cos^{2}\phi\,
Q_{i}\right)\\
&\Om{13}=\frac{3\g}{2}\left(P_{r}-\sin^{2}\theta\cos^{2}\phi\,
Q_{r}\right)\\
&\Ga{23}=\frac{3\g}{2}\left(P_{i}-\sin^{2}\theta\sin^{2}\phi\,Q_{i}\right)\\
&\Om{23}=\frac{3\g}{2}\left(P_{r}-\sin^{2}\theta\sin^{2}\phi\,Q_{r}\right)\\
&\Ga{vc}=-\frac{3\g}{2}\sin^{2}\theta\sin\phi\cos\phi\,Q_{i}\\
&\Om{vc}=-\frac{3\g}{2}\sin^{2}\theta\sin\phi\cos\phi\,Q_{r}
\end{split}\label{cc}
\end{equation}
where for $\xi=R\,\omega_{0}/c$
\begin{equation}
\begin{split}
&P_{i}=\frac{\sin\xi}{\xi}+\frac{\cos\xi}{\xi^{2}}-\frac{\sin\xi}{\xi^{3}},\quad
Q_{i}=\frac{\sin\xi}{\xi}+3\frac{\cos\xi}{\xi^{2}}-3\frac{\sin\xi}{\xi^{3}}\\[2mm]
&P_{r}=\frac{\cos\xi}{\xi}-\frac{\sin\xi}{\xi^{2}}-\frac{\cos\xi}{\xi^{3}},\quad
Q_{r}=\frac{\cos\xi}{\xi}-3\frac{\sin\xi}{\xi^{2}}-3\frac{\cos\xi}{\xi^{3}}
\end{split}\label{PQ}
\end{equation}
From the formulas (\ref{cc}) and (\ref{PQ}) it follows that the
coupling coefficients are small for large distance between the atoms
and tend to zero for $R\to \infty$. On the other hand, when $R\to
0$, $\Om{13},\, \Om{23}$ and $\Om{vc}$ diverge, whereas
$$
\Ga{13},\, \Ga{23}\to \g\quad\text{and}\quad \Ga{vc}\to 0
$$
In the following we will consider two special configurations of
atomic system.
\par
\textit{Configuration I}:  $\theta=\pi$ i.e. both atoms lie along
the $z$ axis and $\phi=\pi/4$. In that case
$$
\Ga{13}=\Ga{23},\quad \Om{13}=\Om{23}
$$
and the coherence terms $\Ga{vc}=\Om{vc}=0$.
\par
\textit{Configuration II}:  $\theta=\pi/2$ i.e. both atoms lie on
the $xy$ plane and $\phi=\pi/4$. In that case
$$
\Ga{13}=\Ga{23},\quad \Om{13}=\Om{23}
$$
and the coherence terms
$$
\Ga{vc}\neq 0,\quad \Om{vc}\neq 0.
$$
\par
The time  evolution of the initial state of two - atomic system is
given by the semi - group $\{ T_{t} \}_{t\geq 0}$ of completely
positive mappings acting on density matrices, generated by
$L^{A}+L^{B}+L^{AB}$. The properties of this semi - group crucially
depend on the distance between the two atoms and the geometry of the
system. It can be shown by a direct calculation, that irrespective
to the geometry, when the distance is large (compared to the
radiation wavelength $\lambda$), the semi - group $\{ T_{t}
\}_{t\geq 0}$ is uniquely relaxing with the asymptotic state
$\ket{3_{A}}\otimes\ket{3_{B}}$. Thus, for any initial state, its
entanglement approaches $0$ when $t\to \infty$. But still there can
be some transient entanglement between the atoms. In the following,
we study in details time evolution of some classes of initial states
and show how the creation of entanglement as well as the process of
disentanglement are sensitive to the geometry of the system.
\section{Negativity}
To describe the process of creation or destruction of entanglement
between the atoms, we need the effective measure of mixed - state
entanglement. For such a measure we take a computable measure of
entanglement proposed in \cite{VW}. The measure is based on the
trace norm of the partial transposition $\ro^{PT}$ of the state
$\ro$. From the Peres - Horodecki criterion of separability \cite{P,
HHH}, it follows that if $\ro^{PT}$ is not positive, then $\rho$ is
entangled and one defines the \textit{negativity} of the state
$\rho$ as
\begin{equation}
N(\ro)=\frac{||\ro^{PT}||-1}{2}
\end{equation}
$N(\ro)$ is equal to the absolute value of the sum of the negative
eigenvalues of $\ro^{PT}$ and is an entanglement monotone, but it
cannot detect bound entangled states \cite{H}.
\par
Although negativity of a given state is easy to compute numerically,
the analytical formulas for general mixed states of two qutrits can
be only obtained for some limited classes of states. The density
matrix $\ro$ which we consider to compute negativity is defined on
the space $\C^{3}\otimes \C^{3}$ and $\ro$ is written in the basis
of product states
\begin{equation}
\ket{j_{A}}\otimes\ket{k_{B}},\quad j,k=1,2,3 \label{basis}
\end{equation}
taken in the lexicographic order. In particular, for the states of
the form
\begin{equation}
\ro=\begin{pmatrix} 0&0&0&0&0&0&0&0&0\\
0&0&0&0&0&0&0&0&0\\
0&0&\ro_{33}&0&0&\ro_{36}&\ro_{37}&\ro_{38}&0\\
0&0&0&0&0&0&0&0&0\\
0&0&0&0&0&0&0&0&0\\
0&0&\ro_{63}&0&0&\ro_{66}&\ro_{67}&\ro_{68}&0\\
0&0&\ro_{73}&0&0&\ro_{76}&\ro_{77}&\ro_{78}&0\\
0&0&\ro_{83}&0&0&\ro_{86}&\ro_{87}&\ro_{88}&0\\
0&0&0&0&0&0&0&0&\ro_{99}
\end{pmatrix}\label{state13}
\end{equation}
the negativity is given by
\begin{equation}
N(\ro)=\frac{1}{2}\left[\,\sqrt{4\,(\,|\ro_{37}|^{2}+|\ro_{38}|^{2}+|\ro_{67}|^{2}+
|\ro_{68}|^{2}\,)+\ro_{99}^{2}}-\ro_{99}\,\right] \label{neg13}
\end{equation}
Notice that (\ref{neg13}) is equal to zero when the coherences
$\ro_{37},\,\ro_{38},\, \ro_{67},\,\ro_{68}$ are all equal to zero,
and is greater then zero when at least one of them is nonzero.
Similarly for the states
\begin{equation}
\ro=\begin{pmatrix} 0&0&0&0&0&0&0&0&0\\
0&\ro_{22}&0&0&0&0&0&0&0\\
0&0&\ro_{33}&0&0&0&\ro_{37}&0&0\\
0&0&0&0&0&0&0&0&0\\
0&0&0&0&0&0&0&0&0\\
0&0&0&0&0&\ro_{66}&0&\ro_{68}&0\\
0&0&\ro_{73}&0&0&0&\ro_{77}&0&0\\
0&0&0&0&0&\ro_{86}&0&\ro_{88}&0\\
0&0&0&0&0&0&0&0&\ro_{99}
\end{pmatrix},\label{state12}
\end{equation}
the negativity can be computed from the formula
\begin{equation}
N(\ro)=\frac{1}{2}\left[\,\sqrt{4\,(\,|\ro_{37}|^{2}+
|\ro_{68}|^{2}\,)+\ro_{99}^{2}}-\ro_{99}\,\right]. \label{neg12}
\end{equation}
\par
On the other hand, for the states
\begin{equation}
\ro=\begin{pmatrix} \ro_{11}&0&0&0&0&0&0&0&0\\
0&0&0&0&0&0&0&0&0\\
0&0&\ro_{33}&0&0&0&\ro_{37}&0&0\\
0&0&0&0&0&0&0&0&0\\
0&0&0&0&0&0&0&0&0\\
0&0&0&0&0&0&0&0&0\\
0&0&\ro_{73}&0&0&0&\ro_{77}&0&0\\
0&0&0&0&0&0&0&0&0\\
0&0&0&0&0&0&0&0&\ro_{99}
\end{pmatrix},\label{state11}
\end{equation}
their negativity
\begin{equation}
N(\ro)=\max\,\left(\,0,\, \widetilde{N}(\ro)\,\right)\label{neg11}
\end{equation}
where
\begin{equation}
\widetilde{N}(\ro)=\frac{1}{2}
\left[\sqrt{(\ro_{11}-\ro_{99})^{2}+4\,|\ro_{37}|^{2}}-\ro_{11}-\ro_{99}\right]
\label{neg11f}
\end{equation}
can be zero, even if the coherence $\ro_{37}$ is not zero. There is
a threshold for the coherence at which two atoms become entangled.
\section{Creation of entanglement}
In this section we study the process of creation of transient
entanglement between atoms prepared in separable initial states. We
fix the distance between the atoms and solve numerically the master
equation (\ref{me}) in two cases of configurations of the system.
\subsection{Initial states $\ket{1_{A}}\otimes\ket{3_{B}}$ and
$\ket{1_{A}}\otimes\ket{2_{B}}$}
\begin{figure}[h]
\centering
{\includegraphics[height=86mm,angle=270]{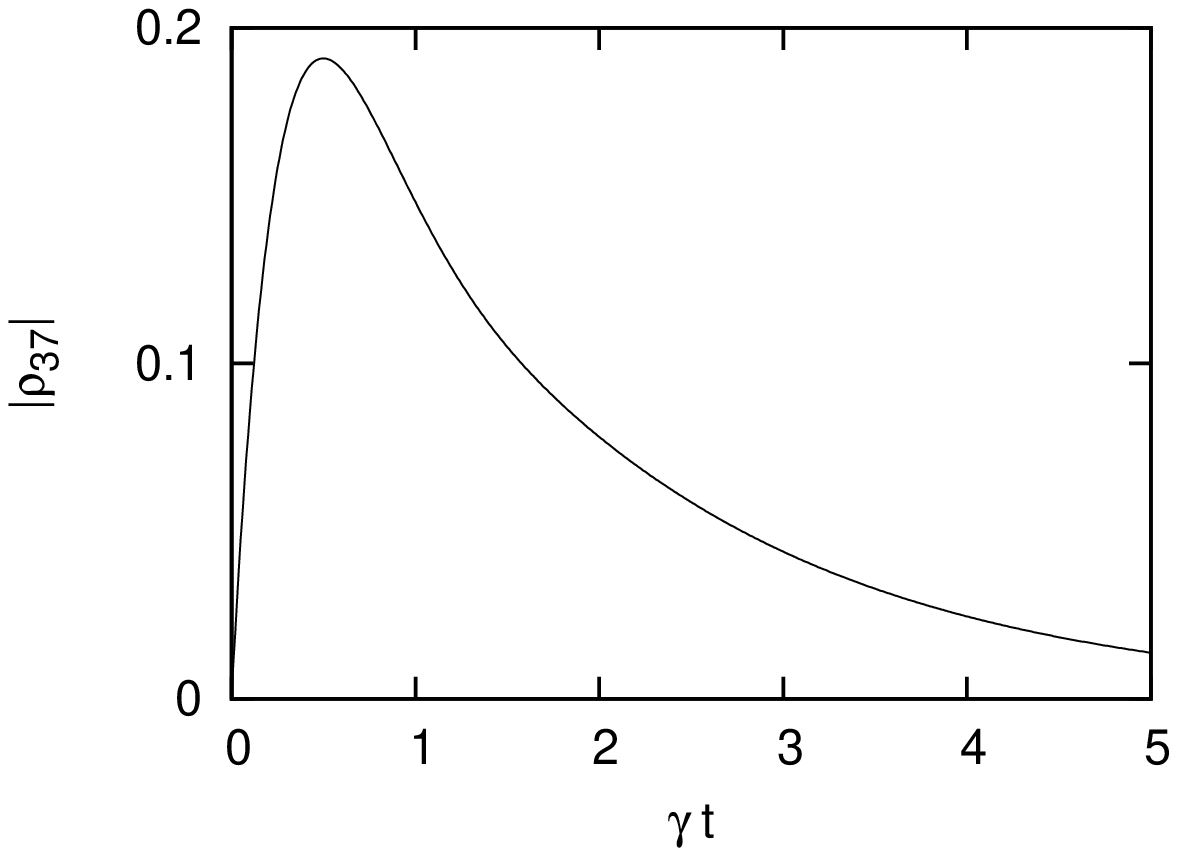}}\caption{The time
evolution of  $|\ro_{37}|$ for the initial state
$\ket{1_{A}}\otimes\ket{3_{B}}$ when $\Ga{vc}=\Om{vc}=0$ and
$R/\lambda=0.2$. }
\end{figure}
When the system is prepared in the pure state $\ket{1_{A}}\otimes
\ket{3_{B}}$ (atom A in the excited state and atom B in the ground
state) and both atoms lie along the $z$ axis (\textit{Configuration
I}), so
$$
\Ga{12}=\Ga{23},\quad \Om{13}=\Om{23}\quad\text{and}\quad
\Ga{vc}=\Om{vc}=0,
$$
one can check that the density matrix at time $t$ takes the form
\begin{equation}
\ro(t)=\begin{pmatrix} 0&0&0&0&0&0&0&0&0\\
0&0&0&0&0&0&0&0&0\\
0&0&\ro_{33}(t)&0&0&0&\ro_{37}(t)&0&0\\
0&0&0&0&0&0&0&0&0\\
0&0&0&0&0&0&0&0&0\\
0&0&0&0&0&0&0&0&0\\
0&0&\ro_{73}(t)&0&0&0&\ro_{77}(t)&0&0\\
0&0&0&0&0&0&0&0&0\\
0&0&0&0&0&0&0&0&\ro_{99}(t)
\end{pmatrix}\label{ro13}
\end{equation}
and by (\ref{neg13})
\begin{equation}
N(t)=\frac{1}{2}\,\left[\,\sqrt{4\,|\ro_{37}(t)|^{2}+\ro_{99}(t)^{2}}-\ro_{99}(t)\,\right]
\label{neg13t}
\end{equation}
Since the process of the photon exchange produces coherence between
the states $\ket{1_{A}}\otimes \ket{3_{B}}$ and $\ket{3_{A}}\otimes
\ket{1_{B}}$, the value of $|\ro_{37}|$ starts to grow and the
system becomes entangled (FIG. 2).
\begin{figure}[h]
\centering
{\includegraphics[height=80mm,angle=270]{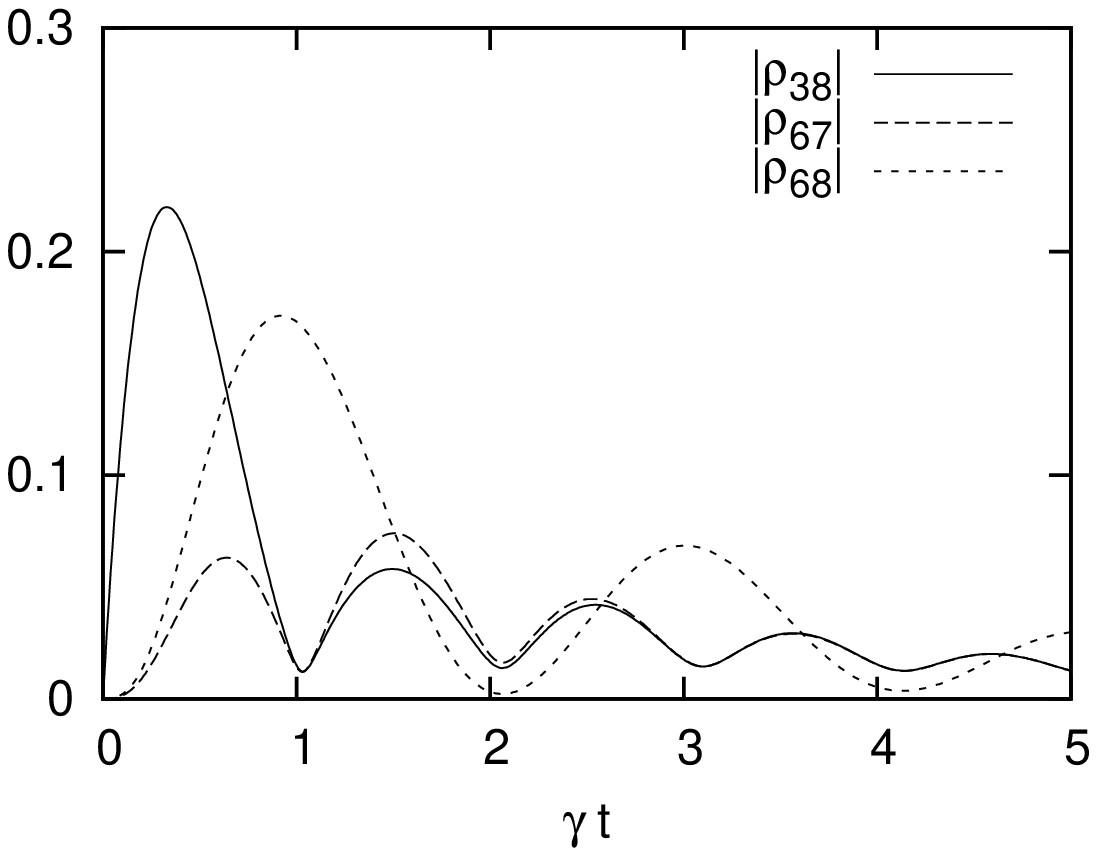}}\caption{The time
evolution of   $|\ro_{38}|$, $|\ro_{67}|$, $|\ro_{68}|$  for the
initial state $\ket{1_{A}}\otimes\ket{3_{B}}$ in the case when
$\Ga{vc}\neq 0,\, \Om{vc}\neq 0$ and $R/\lambda=0.2$. }
\end{figure}
\par
When both atoms lie on the $xy$ plane (\textit{Configuration II}),
the cross coupling coefficients $\Ga{vc}$ and $\Om{vc}$ are nonzero
and the dynamics of the system is changed significantly. The
additional coupling between orthogonal dipoles produces new
coherences $\ro_{36},\, \ro_{38},\, \ro_{67},\, \ro_{68}$ and
$\ro_{78}$, so the state at time $t$ has the form (\ref{state13}).
In particular, the values of $|\ro_{37}|,\, |\ro_{38}|,\,
|\ro_{67}|$ and $|\ro_{68}|$ become nonzero (FIG. 3), so the
negativity of the state can be computed from the formula
(\ref{neg13}).
\begin{figure}[h]
\centering
{\includegraphics[height=86mm,angle=270]{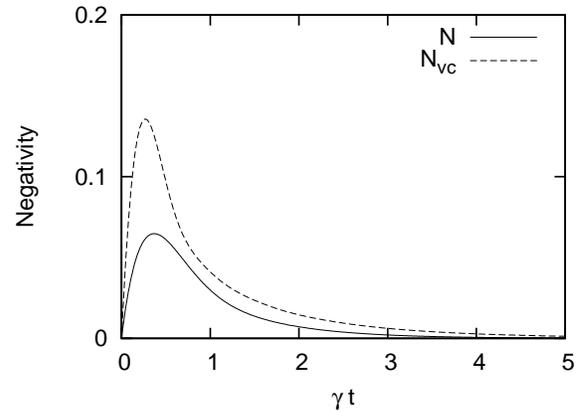}}\caption{The time
evolution of   negativity of initial state
$\ket{1_{A}}\otimes\ket{3_{B}}$ in two cases: $\Ga{vc}\neq 0,\,
\Om{vc}\neq 0$ ($N_{vc}$) and $\Ga{vc}=\Om{vc}=0$ ($N$). In both
cases $R/\lambda =0.2$.}
\end{figure}
In FIG. 4 we plot the time evolution of negativity of initial state
$\ket{1_{A}}\otimes\ket{3_{B}}$ in both configurations. As we see,
the cross coupling between the atoms enhances the production of
entanglement. The same behaviour of negativity can be observed for
initial state $\ket{2_{A}}\otimes\ket{3_{B}}$.
\par
On the other hand, when the system is prepared in the initial state
$\ket{1_{A}}\otimes\ket{2_{B}}$ (both atoms in excited states) and
the cross coupling is absent, the entanglement production is due to
the coherences $\ro_{37}$ and $\ro_{68}$. As in the previous case,
the presence of cross coupling enhances the production of
entanglement, but the maximal value of negativity is much less then
in the case of initial state $\ket{1_{A}}\otimes \ket{3_{B}}$ (FIG.
5).
\begin{figure}[h]
\centering
{\includegraphics[height=85mm,angle=270]{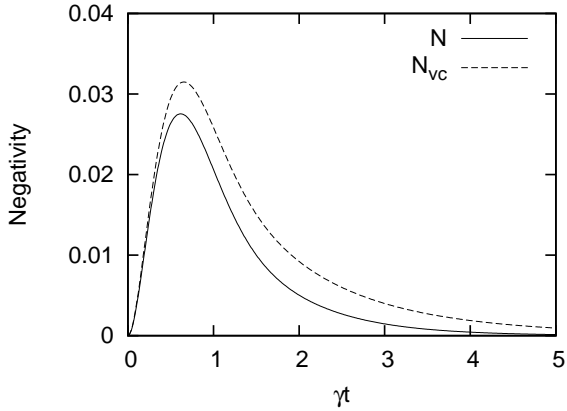}}\caption{The time
evolution of   negativity of initial state
$\ket{1_{A}}\otimes\ket{2_{B}}$ in two cases: $\Ga{vc}\neq 0,\,
\Om{vc}\neq 0$ ($N_{vc}$) and $\Ga{vc}=\Om{vc}=0$ ($N$). In both
cases $R/\lambda=0.2$. }
\end{figure}
\subsection{Initial state $\ket{1_{A}}\otimes\ket{1_{B}}$ and
delayed sudden birth of entanglement}
\begin{figure}[h]
\centering
{\includegraphics[height=88mm,angle=270]{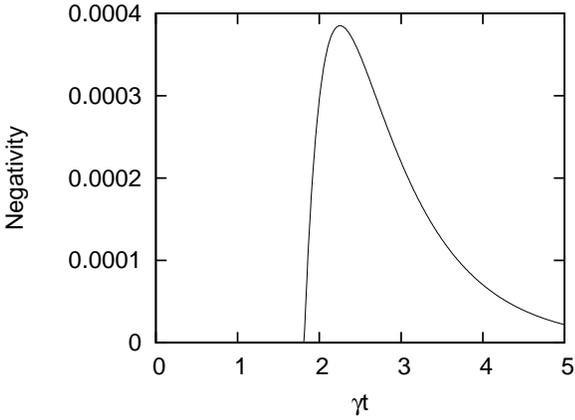}}\caption{The time
evolution of   negativity of initial state
$\ket{1_{A}}\otimes\ket{1_{B}}$. Here we take $\Ga{vc}=\Om{vc}=0$
and $R/\lambda =0.2$.}
\end{figure}
If the system is prepared in the state
$\ket{1_{A}}\otimes\ket{1_{B}}$ (both  atoms are in the same excited
state), and the cross coupling is absent, the state at time $t$
takes the form (\ref{state11}). As in the case of initial state
$\ket{1_{A}}\otimes\ket{3_{B}}$, the entanglement production is due
to the creation of coherence $\ro_{37}$, but in the present case,
the nonzero coherence is only the necessary condition for
entanglement. As it follows from (\ref{neg11}) and (\ref{neg11f}),
there is a threshold for $|\ro_{37}|$ at which the negativity
becomes nonzero. A detailed numerical analysis shows that there is
no entanglement at earlier times, and suddenly at some time the
entanglement starts to build up (FIG. 6). This is the example of
phenomenon of \textit{delayed sudden birth of entanglement}, studied
 by Ficek and Tana\'s \cite{FT} in the case of two - level atoms.
 \begin{figure}[h]
\centering
{\includegraphics[height=85mm,angle=270]{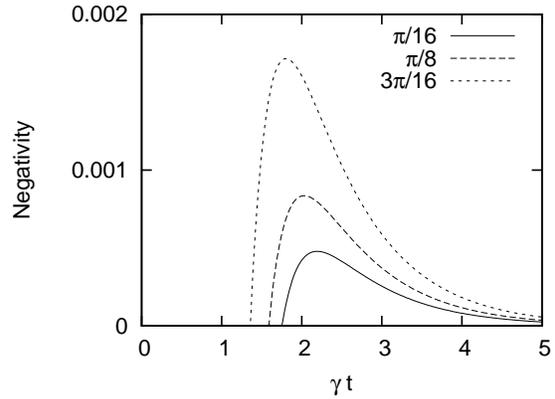}}\caption{The time
evolution of   negativity of initial state (\ref{superposition}) for
different values of $\phi$ ($\Ga{vc}=\Om{vc}=0,\, R/\lambda=0.2$). }
\end{figure}
\begin{figure}[b]
\centering
{\includegraphics[height=80mm,angle=270]{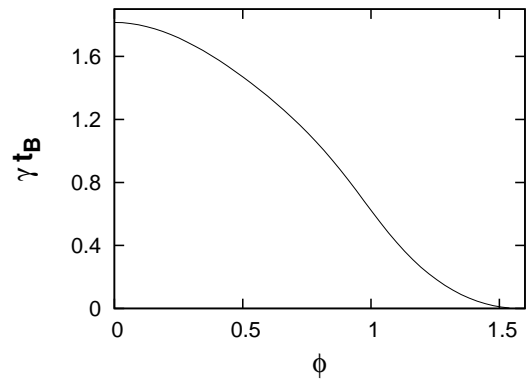}}\caption{The time
of birth of entanglement as a function of $\phi$.}
\end{figure}
To get some insight into the process of creation of entanglement in
this case, consider the initial state
\begin{equation}
\ket{\Psi}=\cos\phi\, \ket{1_{A}}\otimes\ket{1_{B}}+\sin\phi\,
\ket{1_{A}}\otimes\ket{3_{B}},\quad
\phi\in[0,\pi/2]\label{superposition}
\end{equation}
FIG. 7 shows that the evolution of initial state
(\ref{superposition}) crucially depends on the superposition angle
$\phi$.  The smaller is the probability that the system is prepared
in the state $\ket{1_{A}}\otimes\ket{1_{B}}$, the earlier the atoms
become entangled. In FIG. 8 we plot the time of the birth of
entanglement as the function of the superposition angle. This time
is maximal for $\phi=0$ (i.e.
$\ket{\Psi}=\ket{1_{A}}\otimes\ket{1_{B}}$) and is equal to zero for
$\phi=\pi/2$.
\begin{figure}[h]
\centering
{\includegraphics[height=85mm,angle=270]{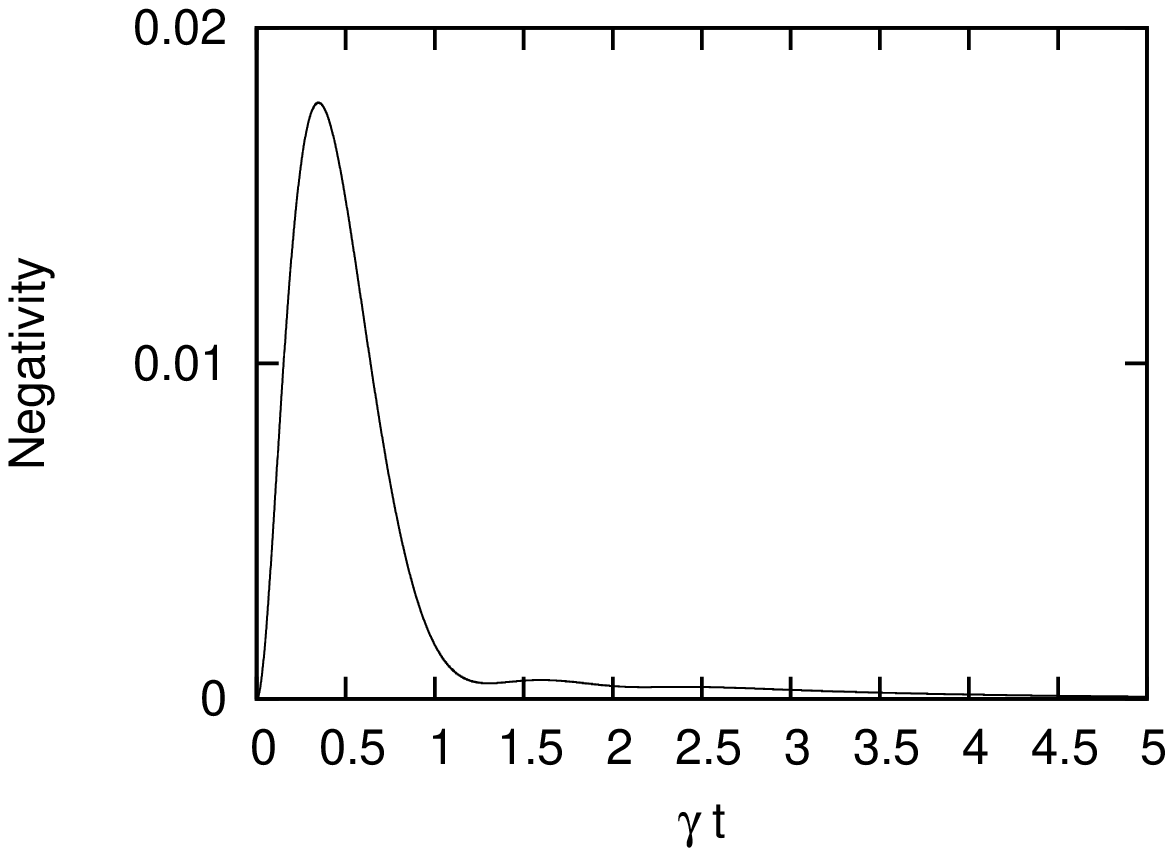}}\caption{The time
evolution of   negativity of initial state
$\ket{1_{A}}\otimes\ket{1_{B}}$ when $\Ga{vc}\neq 0,\, \Om{vc}\neq
0$ and $R/\lambda=0.2$.}
\end{figure}
\par
When the cross coupling coefficients $\Ga{vc}$ and $\Om{vc}$ are not
zero the numerical analysis shows that the time delayed creation of
entanglement does not occur. In that case, even for the initial
state $\ket{1_{A}}\otimes\ket{1_{B}}$ (or
$\ket{2_{A}}\otimes\ket{2_{B}}$), the entanglement starts to build
up immediately after the atoms begin to interact with the vacuum
(FIG. 9).
\section{Disentanglement}
Apart from the effect of creation of entanglement, the quantum
evolution given by the master equation (\ref{me}) may cause also
destruction of correlations, resulting in disentanglement of
initially entangled states. In this section we study the process of
disentanglement for some  entangled pure initial states. We start
with the characterization of maximally entangled states of two three
- level systems (two qutrits).
\subsection{Maximally entangled states of two qutrits and generalized Dicke states}
The basis $\ket{\Psi_{\alpha}},\, \alpha =1,\ldots, 9$ of the space
$\C^{3}\otimes \C^{3}$ consisting of maximally entangled Bell - like
states was constructed in \cite{BHN} (see also \cite{DJ}). The
states $\ket{\Psi_{\alpha}}$ can be written as follows:
\begin{equation}
\begin{split}
&\ket{\Psi_{1}}=\frac{1}{\sqrt{3}}\,\left(\ket{1_{A}}\otimes\ket{1_{B}}
+\ket{2_{A}}\otimes\ket{2_{B}}+\ket{3_{A}}\otimes\ket{3_{B}}\,\right)\\
&\ket{\Psi_{2}}=\frac{1}{\sqrt{3}}\,\left(\ket{1_{A}}\otimes\ket{2_{B}}+
\ket{2_{A}}\otimes\ket{3_{B}}+\ket{3_{A}}\otimes\ket{1_{B}}\,\right)\\
&\ket{\Psi_{3}}=\frac{1}{\sqrt{3}}\,\left(\ket{1_{A}}\otimes\ket{3_{B}}
+\ket{2_{A}}\otimes\ket{1_{B}}+\ket{3_{A}}\otimes\ket{2_{B}}\,\right)\\
&\ket{\Psi_{4}}=\frac{1}{\sqrt{3}}\,\left(\ket{1_{A}}\otimes\ket{1_{B}}+
w\ket{2_{A}}\otimes\ket{2_{B}}+\conj{w}\ket{3_{A}}\otimes\ket{3_{B}}\,\right)\\
&\ket{\Psi_{5}}=\frac{1}{\sqrt{3}}\,\left(\ket{1_{A}}\otimes\ket{2_{B}}+
w\ket{2_{A}}\otimes\ket{3_{B}}+\conj{w}\ket{3_{A}}\otimes\ket{1_{B}}\,\right)\\
&\ket{\Psi_{6}}=\frac{1}{\sqrt{3}}\,\left(\ket{1_{A}}\otimes\ket{3_{B}}+
w\ket{2_{A}}\otimes\ket{1_{B}}+\conj{w}\ket{3_{A}}\otimes\ket{2_{B}}\,\right)\\
&\ket{\Psi_{7}}=\frac{1}{\sqrt{3}}\,\left(\ket{1_{A}}\otimes\ket{1_{B}}+
\conj{w}\ket{2_{A}}\otimes\ket{2_{B}}+w\ket{3_{A}}\otimes\ket{3_{B}}\,\right)\\
&\ket{\Psi_{8}}=\frac{1}{\sqrt{3}}\,\left(\ket{1_{A}}\otimes\ket{2_{B}}+
\conj{w}\ket{2_{A}}\otimes\ket{3_{B}}+w\ket{3_{A}}\otimes\ket{1_{B}}\,\right)\\
&\ket{\Psi_{9}}=\frac{1}{\sqrt{3}}\,\left(\ket{1_{A}}\otimes\ket{3_{B}}+
\conj{w}\ket{2_{A}}\otimes\ket{1_{B}}+w\ket{3_{A}}\otimes\ket{2_{B}}\,\right)
\end{split}\label{bell}
\end{equation}
where
$$
w=e^{2\pi i/3}
$$
One can check that the states (\ref{bell}) have maximal negativity
and that they are locally equivalent.
\par
There is another class of pure entangled states of two qutrits which
are very important for the analysis of the dynamics of coupled three
- level atoms. The \textit{generalized symmetric and antisymmetric
Dicke states} (see e.g. \cite{BGZ}), defined by the formulas
\begin{equation}
\begin{split}
&\ket{s_{kl}}=\frac{1}{\sqrt{2}}\,\left(\ket{k_{A}}\otimes\ket{l_{B}}
+\ket{l_{A}}\otimes\ket{k_{B}}\,\right)\\
&\ket{a_{kl}}=\frac{1}{\sqrt{2}}\,\left(\ket{k_{A}}\otimes\ket{l_{B}}
-\ket{l_{A}}\otimes\ket{k_{B}}\,\right)
\end{split}\label{dicke}
\end{equation}
where $k,l=1,2,3;\, k<l$, are not maximally entangled (their
negativity is equal to $1/2$) but have a remarkable properties. As
it was shown in our previous paper \cite{DJ1}, in the limit of small
separation between the atoms, the process of photon exchange
produces such correlations that the dynamics is not ergodic and
there are nontrivial asymptotic stationary states. In that case, the
symmetric Dicke states $\ket{s_{kl}}$ decay to the ground state
$\ket{3_{A}}\otimes\ket{3_{B}}$ whereas antisymmetric states
$\ket{a_{13}}$ and $\ket{a_{23}}$ decouple from the environment and
therefore are stable. Moreover, the state $\ket{a_{12}}$ is not
stable, but is asymptotically nontrivial. So for the distances
comparable to the radiation wavelength $\lambda$, symmetric and
antisymmetric Dicke states will decay with significantly different
rates and the populations in the antisymmetric states can be used to
characterize disentanglement properties of given initial state.
\subsection{Time evolution of Dicke states}
Now we consider the evolution of the antisymmetric Dicke state
\begin{equation}
\ket{a_{13}}=\frac{1}{\sqrt{2}}\,\left(\ket{1_{A}}\otimes\ket{3_{B}}-
\ket{3_{A}}\otimes\ket{1_{B}}\,\right)\label{a13}
\end{equation}
in the case when the  distance between the atoms is comparable to
$\lambda$. If $\Ga{vc}=\Om{vc}=0$, the state at time $t$ takes the
form (\ref{ro13}), so the degree of its entanglement is determined
by the coherence $\ro_{37}(t)$. The same is true for the symmetric
state
\begin{equation}
\ket{s_{13}}=\frac{1}{\sqrt{2}}\,\left(\ket{1_{A}}\otimes\ket{3_{B}}+
\ket{3_{A}}\otimes\ket{1_{B}}\,\right)\label{s13}
\end{equation}
which decays to the ground state even in the limit of small
separation. As we show numerically, time evolution of $\ro_{37}$ for
the symmetric state (\ref{s13}),  differs significantly from that
for antisymmetric state (\ref{a13}) (see FIG. 10) and the latter
disentangle much slower that the former (FIG. 11).
\begin{figure}[h]
\centering
{\includegraphics[height=85mm,angle=270]{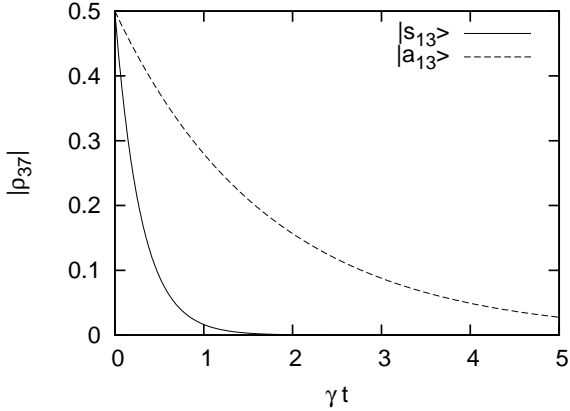}}\caption{The
time evolution of $|\ro_{37}|$ for the symmetric  and antisymmetric
 Dicke states in the case $\Ga{vc}=\Om{vc}=0$ and $R/\lambda=0.2
$.}
\end{figure}
\begin{figure}[h]
\centering
{\includegraphics[height=85mm,angle=270]{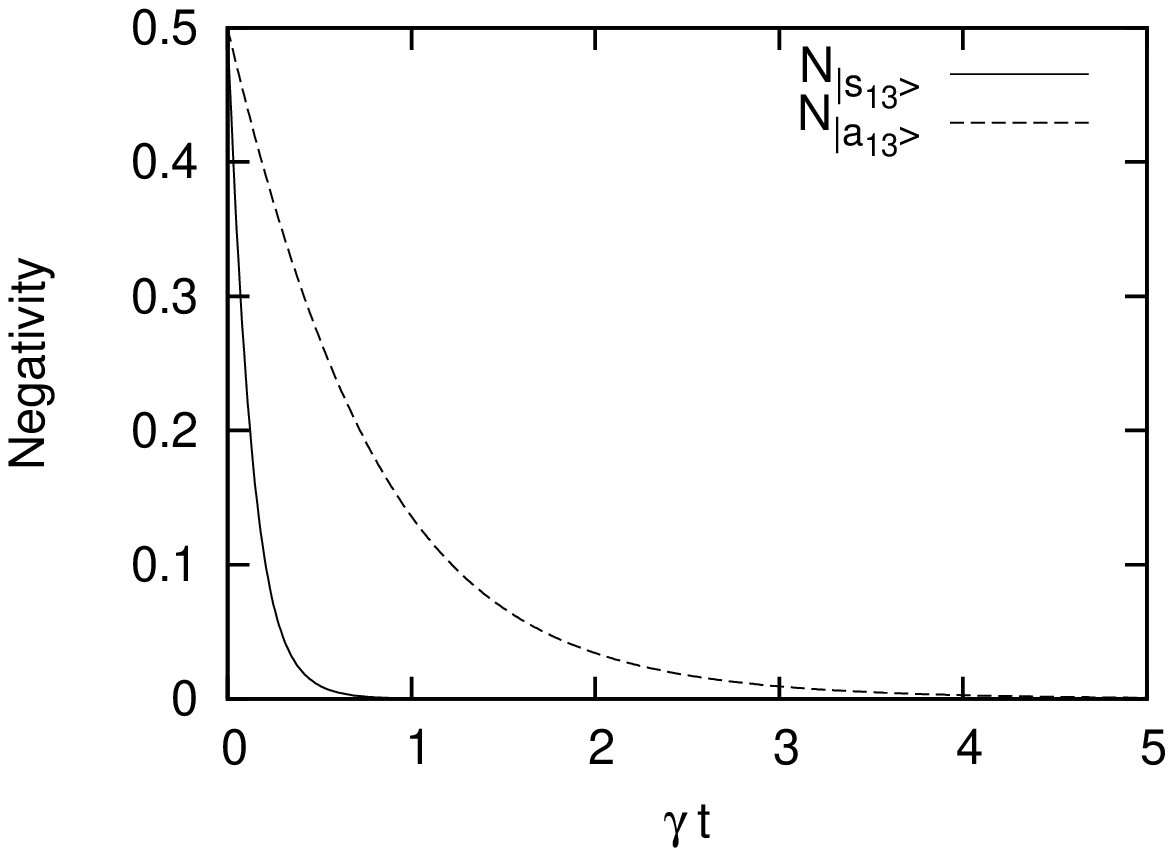}}\caption{Disentanglement
of Dicke states $\ket{s_{13}}$  and $\ket{a_{13}}$
($\Ga{vc}=\Om{vc}=0,\, R/\lambda=0.2$).}
\end{figure}
\par
When the cross coupling coefficients are not zero, the dynamics of
Dicke states is more complicated, and at a given time $t$  they have
the form (\ref{state13}). Detailed analysis of the evolution of
negativity indicates that for the antisymmetric state
$\ket{a_{13}}$, additional coupling between transitions with
orthogonal dipole moments, slow down the process of disentanglement
(FIG. 12). On the other hand, this coupling does not influence rapid
disentanglement of the symmetric state.
\par
The antisymmetric state $\ket{a_{12}}$ is not stable in the limit of
small separation between the atoms, but it is asymptotically
nontrivial \cite{DJ1}. It can be shown that $\ket{a_{12}}$ evolves
to the asymptotic state $\ro$ which has the form (\ref{state12})
with
$$
\ro_{22}=\ro_{99}=0,\quad
\ro_{33}=\ro_{66}=\ro_{88}=\frac{1}{4},\quad
\ro_{37}=\ro_{68}=-\frac{1}{4}
$$
so by (\ref{neg12}), the asymptotic negativity of $\ket{a_{12}}$ has
the value $\sqrt{2}/4$. For the atom separation comparable with
$\lambda$, this state disentangle quicker then $\ket{a_{13}}$.
\begin{figure}[h]
\centering
{\includegraphics[height=85mm,angle=270]{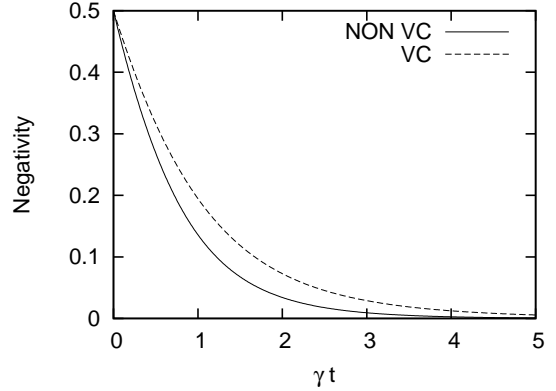}}\caption{Disentanglement
of Dicke state $\ket{a_{13}}$ in two cases: $\Ga{vc},\, \Om{vc}\neq
0 $ (VC) and $\Ga{vc}=\Om{vc}=0$ (NON VC). In both cases $R/\lambda
=0.2$}
\end{figure}
\subsection{Disentanglement of Bell states}
As it was stated before, in the case of small separation between the
atoms, the antisymmetric Dicke states $\ket{a_{12}},\, \ket{a_{23}}$
are stable, and the state $\ket{a_{12}}$ has a nontrivial asymptotic
limit. For that reason, the initial states which have the property
of trapping populations in $\ket{a_{13}},\, \ket{a_{23}}$  and
$\ket{a_{12}}$, decay to an entangled asymptotic states. In the
process of evolution, the initial populations in $\ket{a_{13}}$ and
$\ket{a_{23}}$ are conserved, whereas the population in
$\ket{a_{12}}$ can be transformed into $\ket{a_{13}}$ and
$\ket{a_{23}}$, giving the enlargement of initial populations.
\par
Consider now the Bell states (\ref{bell}). One checks that the
states $\ket{\Psi_{1}},\, \ket{\Psi_{4}}$ and $\ket{\Psi_{7}}$ have
zero populations in $\ket{a_{13}},\, \ket{a_{23}}$ and
$\ket{a_{12}}$, so they decay to the separable asymptotic state. In
fact, the limiting state in this case is the ground state
$\ket{3_{A}}\otimes\ket{3_{B}}$. On the other hand, the remaining
Bell states $\ket{\Psi_{2}},\, \ket{\Psi_{3}},\, \ket{\Psi_{5}},\,
\ket{\Psi_{6}},\, \ket{\Psi_{8}}$ and $\ket{\Psi_{9}}$ have equal
populations in antisymmetric Dicke states, so they have the same
asymptotic entanglement. Take for example the state
$\ket{\Psi_{2}}$. Since the corresponding populations are equal
$1/6$, by the general result of \cite{DJ1} the asymptotic state is
of the form
\begin{equation}
\ras=\begin{pmatrix} 0&0&\hspace*{2mm}0&0&0&\hspace*{2mm}0&\hspace*{2mm}0
&\hspace*{2mm}0&\hspace*{2mm}0\\[2mm]
0&0&\hspace*{2mm}0&0&0&\hspace*{2mm}0&\hspace*{2mm}0&\hspace*{2mm}0&\hspace*{2mm}0\\[2mm]
0&0&\hspace*{2mm}\frac{1}{8}&0&0&-\frac{1}{12}&-\frac{1}{8}
&\hspace*{2mm}\frac{1}{12}&\hspace*{2mm}\frac{1}{12}\\[2mm]
0&0&\hspace*{2mm}0&0&0&\hspace*{2mm}0&\hspace*{2mm}0
&\hspace*{2mm}0&\hspace*{2mm}0\\[2mm]
0&0&\hspace*{2mm}0&0&0&\hspace*{2mm}0&\hspace*{2mm}0&\hspace*{2mm}0&\hspace*{2mm}0\\[2mm]
0&0&-\frac{1}{12}&0&0&\hspace*{2mm}\frac{1}{8}
&\hspace*{2mm}\frac{1}{12}&-\frac{1}{8}&-\frac{1}{12}\\[2mm]
0&0&-\frac{1}{8}&0&0&\hspace*{2mm}\frac{1}{12}
&\hspace*{2mm}\frac{1}{8}&-\frac{1}{12}&-\frac{1}{12}\\[2mm]
0&0&\hspace*{2mm}\frac{1}{12}&0&0&-\frac{1}{8}
&-\frac{1}{12}&\hspace*{2mm}\frac{1}{8}&\hspace*{2mm}\frac{1}{12}\\[2mm]
0&0&\hspace*{2mm}\frac{1}{12}&0&0&-\frac{1}{12}
&-\frac{1}{12}&\hspace*{2mm}\frac{1}{12}&\hspace*{2mm}\frac{1}{2}
\end{pmatrix}
\end{equation}
Its negativity can be computed numerically and one obtains that
$N(\ras)\simeq 0.0968$.
\par
If the distance $R$ is comparable with the wavelength $\lambda$ and
the cross coupling is absent, the populations in $\ket{a_{13}}$ and
$\ket{a_{23}}$ are no longer conserved but increase at the beginning
(since the population in $\ket{a_{12}}$ decreases) and then decay
much slower than the populations of the remaining states (FIG. 13).
\begin{figure}[h]
\centering
{\includegraphics[height=85mm,angle=270]{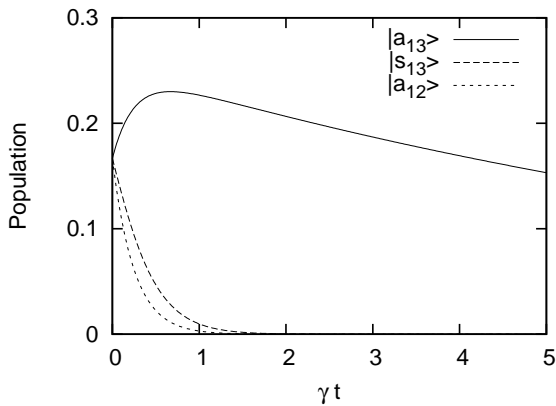}}\caption{Time
evolution of population in antisymmetric state $\ket{a_{13}}$,
symmetric state $\ket{s_{13}}$ and antisymmetric state
$\ket{a_{12}}$  for the initial Bell state $\ket{\Psi_{2}}$. We take
here $\Ga{vc}=\Om{vc}=0$ and $R/\lambda =0.08$.}
\end{figure}
In this way, the entanglement of state $\ket{\Psi_{2}}$ is more
robust against the noise than the entanglement of $\ket{\Psi_{1}}$
(FIG. 14).
\begin{figure}[h]
\centering
{\includegraphics[height=85mm,angle=270]{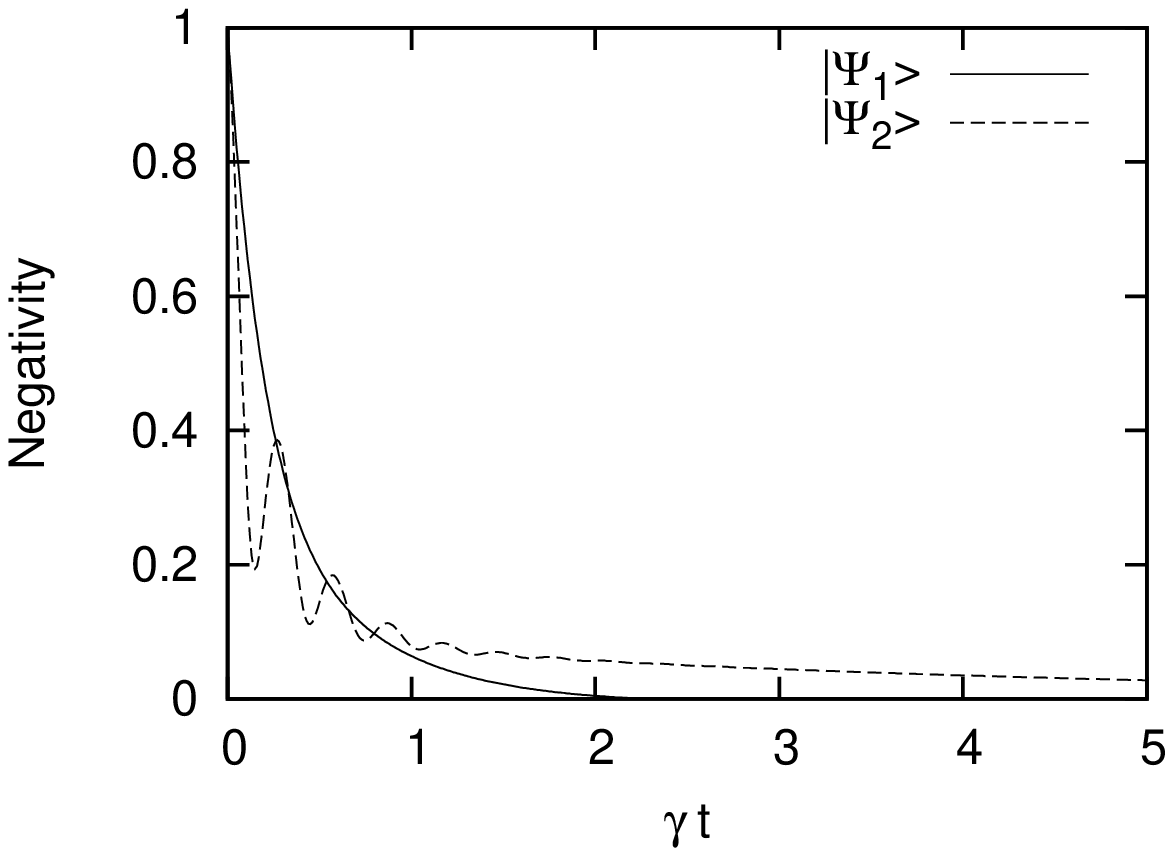}}\caption{Disentanglement
of the Bell state $\ket{\Psi_{1}}$  and $\ket{\Psi_{2}}$ . We take
here $\Ga{vc}=\Om{vc}=0$ and $R/\lambda=0.08$.}
\end{figure}
The same is true for the other states from this class. So all Bell
states can be divided into two classes. The states from the first
class containing $\ket{\Psi_{1}},\, \ket{\Psi_{4}}$ and
$\ket{\Psi_{7}}$ decay rapidly, whereas remaining states decay much
slower. Since the Bell states are locally equivalent, local
operations performed on the initial state can change the robustness
of  entanglement against the noise. On the other hand, as we show
numerically, the influence of cross coupling between the atoms on
the process of disentanglement of Bell states is negligible (FIG.
15).
\begin{figure}[ht]
\centering
{\includegraphics[height=85mm,angle=270]{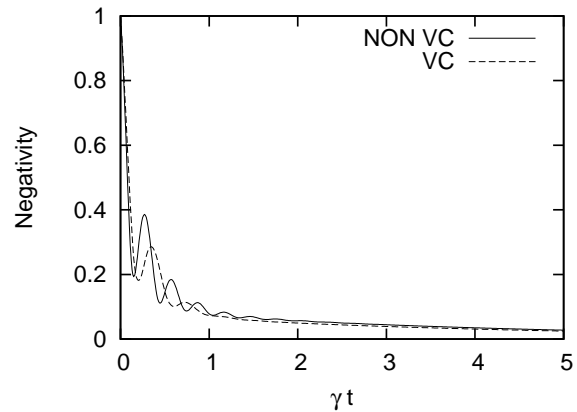}}\caption{Disentanglement
of the Bell state $\ket{\Psi_{2}}$ in two cases: $\Ga{vc}=\Om{vc}=0$
(NON VC) and $\Ga{vc},\, \Om{vc}\neq 0$(VC). In both cases
$R/\lambda=0.08$.}
\end{figure}
\section{Conclusions}
We have studied the dynamics of entanglement in the system of three
- level atoms in the \textsf{V}  configuration, coupled to the
common vacuum and separated by a distance comparable to the
radiation wavelength. In  this case only some transient entanglement
between the atoms can exist but the dynamical generation of such
entanglement is possible. It happens for example, when the cross
coupling between orthogonal dipoles is absent and initially only one
atom is excited. Additional coupling enhances the production of
entanglement and causes that entanglement can be produced also in
the case when two atoms are excited. Initial states with two atoms
excited lead also to the interesting phenomenon of delayed sudden
birth of entanglement. The process of disentanglement of initially
entangled states is less sensitive to cross coupling between the
atoms. We have shown this for the maximally entangled Bell states.
On the other hand, the rate of disentanglement of Bell states
crucially depends on populations of initial state in the
antisymmetric Dicke states, which are more robust against the noise
then the Bell states. We have demonstrated that those Bell states
which have no populations in the antisymmetric Dicke states rapidly
disentangle, whereas remaining Bell states disentangle much slower.

\end{document}